\documentclass[12pt,preprint]{aastex}

\def\ro{{\it ROSAT\/}}
\def\asca{{\it ASCA\/}}
\def\xmm{{\it XMM-Newton\/}}

\shorttitle{A refined ephemeris and spectroscopy of Geminga}
\shortauthors{Jackson and Halpern}

\begin{document}

\title{A refined ephemeris and phase resolved X-ray spectroscopy of the Geminga pulsar}

\author{M. S. Jackson and J. P. Halpern}
\affil{Department of Astronomy, Columbia University, New York, NY, 10025-6601}

\begin{abstract}
We present a refined phase-connected post-glitch ephemeris for the 
Geminga pulsar that is a good fit to all the post-glitch data from EGRET, 
\asca, and \xmm. We also present the results of phase-resolved 
spectroscopy of two \xmm\
X-ray observations of the Geminga pulsar obtained in 2002 and 2004. An investigation is made into a previously claimed existence
of a small hot spot on the neutron star surface.  We conclude that
that interpretation was more likely an artifact of an overly restrictive
assumption used to fit the phase-resolved spectra, namely, that the spectral
index of the non-thermal component is constant.  When we allow the
spectral index to vary as a function of rotation phase, we find systematic
variations in spectral index, and such fits do not require an additional,
hot blackbody component.

\end{abstract}

\keywords{pulsars: individual (Geminga)--- stars: neutron --- X-rays: stars}

\section{Introduction}

Since its discovery in 1972 \citep{fi75,th77}, the Geminga pulsar has
been determined to be a relatively old radio-quiet pulsar with a
period of 237~ms, which modulates in X-rays \citep{hh92}, $\gamma$-rays
\citep{be92}, and at optical wavelengths \citep{sh98} (see \cite{bc96}
for a review). A glitch
occurred in late 1996 and the post-glitch ephemeris was calculated
from EGRET data \citep{mj02}.  The latest \xmm\ observations have been
used to further refine the ephemeris. The method and resulting new
ephemeris are presented here.

Unlike the observed hard X-ray and $\gamma$-ray emission, which is
thought to arise in the magnetosphere, the soft X-ray and ultraviolet
radiation is thermal in nature and is produced at the neutron star
surface itself. As a neutron star of Geminga's age cools, the
interior heat travels preferentially along the magnetic field lines,
producing relatively warm areas at the poles. As the star rotates, the
amount of surface area
of these areas exposed to the observer changes, giving rise to pulsed
emission. The shape of the thermal light curves is strongly determined
by the angle between the magnetic and rotation axes, and by the
position of the observer with respect to the rotation axis.  

The 2002 \xmm\ observation allowed for the first time the opportunity
to perform statistically significant phase-resolved spectroscopy on
Geminga data. Prior to this, the \asca\ data provided for dividing the
spectra into two parts based on the phase, and the difference in the
fitted parameters was only marginally significant \citep{mj02}. With
phase-resolved spectroscopy it is possible to determine the changes in
the individual spectral components as the pulsar rotates, to give a
more thorough picture of the various emission regions. 

It has recently been proposed \citep{ca04,de04} that the pulsed
thermal emission from Geminga and other pulsars is composed
of a power law component and two blackbody components, one arising
from a large warm area on the neutron star surface and the other from
a hot spot located at the polar cap. Those authors performed fits on the
phase-resolved spectra by freezing the hydrogen column density,
warm and hot blackbody temperatures, and photon index at the values
obtained from a fit to the entire spectrum. It is expected that
the column density and temperatures would remain constant, considering
the physical meaning of those parameters. However, there is no reason that 
the photon index, which is a property of the emission from the
magnetosphere, would remain constant with rotation phase. The blackbody 
components represent thermal emission
from the surface of the pulsar, but the power-law component is only a
parameterization of the magnetospheric emission, and it is used as a
convenience for fitting the pulsar spectrum in a particular energy
range, without a rigorous physical model to support its use.

For all three pulsars that were investigated in this manner
\citep{de04}, the authors found a phase difference between the 
warm and hot emission, and in fact for the pulsar PSR0656+14 these two
blackbody components are almost exactly out of phase with each other,
meaning that the hot spot and warm region are not in the same location,
leading to the question of whether there are two separate heating
mechanisms at work, or whether the thermal conductivity on the surface
does not behave as expected. \cite{ka05} bring into question the existence 
of the hot spot on
Geminga because the radius for the hot spot (50 m) is much less
than the expected size of the polar cap at the surface (300
m). However, those authors did not perform phase resolved spectroscopy
and did not offer an alternative model or an explanation for the
excess of counts above the model in some of the X-ray spectra. In their
paper, an attempt is made to connect the spectrum among the visual,
UV, and X-ray ranges. Additional observations would be required at the
unmeasured energy bands to discover the full spectral shape. 

\section{Observations and data preparation}\label{dataprep}

A log of the observations used in this paper is given in Table \ref{tbl-1}.
A long observation of Geminga was performed by \xmm\ in 2002, and a
shorter observation in April 2004, for the primary purpose of
maintaining the phase-connected ephemeris. The shorter 2004 observation
does not provide enough counts to use it exclusively in order to
perform phase-resolved spectroscopy.

The X-ray data used in this paper are primarily the EPIC-pn data from the 
two \xmm\ observations.
The data from the EPIC-MOS instruments do not have sufficient timing
resolution for phase resolved spectroscopy. MOS
spectra can be used to fit the entire spectrum to make
sure the fitted parameters are reasonable.

The EGRET and \asca\ GIS data used are as described in \cite{mj02},
and were extracted and prepared in the same manner as previously
published. The resulting data are in the form of barycenter-corrected
event files, containing positions and arrival times of detected photons.

The latest version of the Science Analysis System (SAS) package
(version 6.1)  for \xmm\ data analysis was used to process the \xmm\ data
into event files. The SAS  package contains a useful ftool called
{\it barycen} that performs the barycentric correction on \xmm\ EPIC-pn
data. It uses many of the housekeeping files that are included with
the data, for spacecraft position, etc. The timing resolution of the
pn instrument in small window mode is 5.7~ms, which is 
sufficient for timing
analysis of Geminga.

\begin{figure}
\scalebox{0.85}{\rotatebox{270}{\plotone{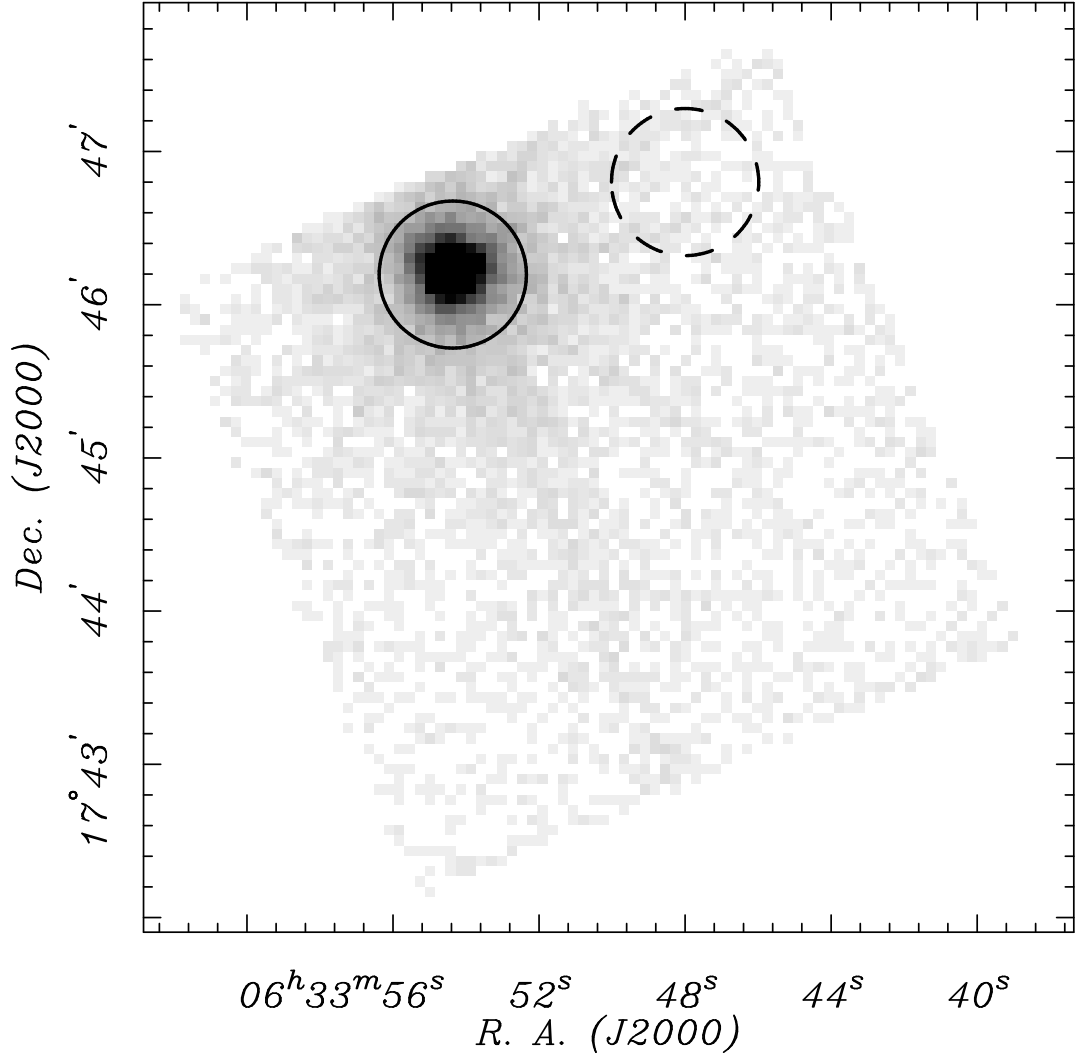}}}
\caption{\xmm\ pn image of Geminga, showing the source circle ({\it solid
  line}) and background circle ({\it dashed line})
from which the background was subtracted for calculation of the
 light curves and spectra. The radii of the circles
($29^{\prime\prime}$) were
chosen to maximize the signal-to-noise ratio in the light curve.
\label{image}}
\end{figure}

The data for all \xmm\ pn observations were extracted from a circle of
radius $29^{\prime\prime}$, with a background circle with the same
radius, offset from the source, as shown in Figure \ref{image}. These
regions were chosen to maximize the signal to noise ratio. Events were
selected for both spectra and light curves based on the FLAG parameter
set to zero, and the PATTERN parameter of 4 or less. There was no
evidence for pileup in either observation. 

For the phase-resolved spectra an additional criterion that
the chosen events fall into phase bins of 0.1 cycles is used, and
one spectrum is produced for each of ten equally spaced phase ranges,
with zero phase corresponding to T$_0$ of the ephemeris given in Table
2. The SAS program {\it especget} extracts source and background
spectra and produces arf and rmf response files. The spectra were
binned with a minimum of 40 counts in each bin for the whole spectrum
and 25 counts for the phase-resolved spectra.

\subsection{XMM Timing errors}

It should be noted that the 2002 \xmm\ observation was found to have a
timing jump of 7 seconds
for approximately one-third of the observing time. This arose from three
of the bit counters getting flipped. It was corrected when the timing
counter reached its maximum and was reset. Before the latest version
of SAS (6.1) was released, it was necessary to correct
the shifted times by hand, and this greatly improved the appearance of
the light curves. Version 6.1 of SAS now corrects for this error,
though it is as yet uncertain if it will correct for all such errors
in future observations.

It is suspected that the timing errors in \xmm\ pn data occur between
the times when the timing counter is reset, which happens at the
beginning of every observation and once or more during long
observations. It was believed that it was not possible for there to be
a timing error in a single short observation, which does not have enough
duration for the timing counter to have reached its maximum and
reset. 

Figure \ref{phaseplot} shows the phase residuals for all X-ray and
EGRET observations of Geminga using the preglitch ephemeris given in
\cite{ma98}. The phase residual of the 2004 \xmm\ observation lies at
a phase that is approximately 0.2 (or 0.8) away from the ephemeris
calculated in this paper from the EGRET points. This indicates either
that a second (comparatively larger than that of 1996) glitch occurred
between the epochs of the 2002
and 2004 \xmm\ observations, or that there was a timing error in the
2004 observation that was not corrected by the software. Future
observations will determine which of these possible explanations is
correct.

If the phase jump resulted from a glitch, an estimate can be made of
the glitch size. With only one post-glitch point there is no way of
estimating the epoch of the glitch, but assuming that the possible
glitch occurred shortly after the 2002 observation, the value of $\Delta
f/f$ is estimated to be $3 \times 10^{-9}$, which is 5 times the size
of the 1996 glitch. That value is a lower limit for the glitch
magnitude. If the possible glitch occurred later, or if the phase
residual is, say, --1.8 instead of --0.8, that would result in
a larger calculated value of the glitch. 

If the phase shift resulted from a timing error, the most likely
correction is to subtract 1 second from each arrival time. Out of all
possible corrections, that of --1 second yielded a result that best
lines up with the calculated ephemeris. The possible corrected point is
indicated by the open circle beneath the 2004 point in Figure
\ref{phaseplot}. 

Whether a second
glitch did occur will be determined from the next Geminga
observation. However, as this data set was only used to check the new
ephemeris and not to calculate it, the timing error (if any) is somewhat
unimportant to the results of this paper, for timing up to epoch of the 2002
\xmm\ observation, or for the spectroscopy.  This timing
issue does not affect the results of the phase resolved spectroscopy,
as the data were lined up in phase with the 2002 observation before
the spectra were extracted.

\section{Determination of the Post-Glitch Ephemeris}

The EGRET observations fortunately provide data for a timeline of
nearly 1400 days after the glitch, which means that from the
EGRET data alone, the post-glitch ephemeris parameters can be
determined. All of the EGRET photon arrival times after the 9th EGRET
observation (approximately when the glitch occurred) were used to
find $f$ and $\dot f$.  

An iterative search has been performed for the $f$ and $\dot f$
parameters, by using  $Z_n^2$ and folding techniques. The search
alternates between finding the best value for $f$ and $\dot f$, while
holding the other fixed.  The
value of $Z_2^2$ is determined at each $f$ or $\dot f$, and light curves are
made with 12 and 20 bins, from which $\chi^2$ is determined
for each. The value of the varying parameter that gives the maximum
$Z_2^2$ and $\chi^2$ is the best, and is then
used for the next iteration. 

To establish initial $f$ and $\dot f$, the first and last post-glitch EGRET
observations are used (the 10th and 14th observations in Table 1), and
a frequency is determined for each with $\dot f$ set at
zero. From these quantities and the amount of elapsed time
between the two observations, a value of $\dot f$ is calculated. This
preliminary $\dot f$ is used in a search for $f$, using all five
post-glitch EGRET observations (10--14 in Table 1), centered at the
preliminary frequency as determined from the first post-glitch
observation. Given this newly determined frequency, a search is
then done for $\dot f$, centered at its preliminary value. This process
is continued iteratively until the parameters don't change from one
iteration to the next. These parameters are then checked with the XMM data.

The uncertainties in the $f$ and $\dot f$ parameters estimated from this
method are given by a reduction in the statistic of approximately 1
$\sigma$, based on the number of degrees of freedom implicit in the
statistic. The uncertainty values from the $Z_2^2$ and folding $\chi^2$
statistics are usually in good agreement.

The resulting post-glitch ephemeris, with an epoch $T_0$ chosen so
that the phase lines up with the previous ephemeris, is given in Table
2, and the summed EGRET light curve folded at the new ephemeris is shown in the bottom panel of
Figure \ref{xraylc}. The post-glitch EGRET observations show a much more consistent position of
the peaks in phase, compared with one another and with the previous
ephemeris. While it is not possible to determine the exact epoch of
the glitch, the extrapolation of the post-glitch points in Figure
\ref{phaseplot} give a good estimation. The glitch occurred at a time
very soon after the 9th EGRET observation in late 1996. By adjusting
the estimated glitch epoch, it is possible to determine the point at
which the curve given by the new ephemeris lines up best with the
post-glitch EGRET, \asca\, and \xmm\ points in Figure
\ref{phaseplot}. This epoch was found to be MJD 50320.

\subsection{Refining the ephemeris using \xmm\ pn data}

Given $f$ and $\dot f$ parameters from the EGRET photon arrival times up to
2000, the newer XMM observations can be used to test the validity
and continuity of these values. Whereas the absolute phases of the EGRET
peaks can be determined by using a fit to Lorentzians as described in
\cite{mj02}, it is not as straightforward to determine the absolute
phases of the X-ray light curves. The phases of the X-ray peaks
relative to the $\gamma$-ray peaks have been established with
previous EGRET and ASCA data, so a similar comparison can be made
between the EGRET peaks and XMM light curves from a more recent
epoch. Figure \ref{xraylc} shows the 1994 and 1999 \asca\ GIS, and
2002 and 2004 \xmm\ pn light curves, along with the EGRET
light curve. As can be seen from that Figure, the X-ray peaks line up
with each other 
using the new ephemeris, and the large peak in the 0.7 -- 2.0 keV X-ray
light curves consistently occurs approximately 0.05 later in phase than
the first EGRET peak. 

To determine the phase difference between two X-ray observations, the
light curves for an identical energy range are compared in a bin by bin
fashion, and a $\chi^2$ value is calculated. This is done while
varying the phase of the second light curve between 0.00 and 0.99 in 0.01
increments. The calculated phase difference between the two data sets
is given by the value that produces the smallest $\chi^2$. 

The uncertainty on the phase for a given X-ray observation is
calculated by creating a template light curve, composed of the 2002 and
2004 \xmm\ EPIC-pn observations, folded at an ephemeris determined by a
method similar to that described in Section 5. This template
light curve is compared to each of the X-ray light curves from the
\asca\ and \xmm\ observations and is shifted away from its best value
until the $\chi^2$ value increases by 1$\sigma$. This gives the
uncertainty for each of the labeled X-ray points shown in Figure
\ref{phaseplot}. An example showing the template light curve compared
with the 1994 \asca\ GIS light curve is shown in Figure \ref{complc}.

\section{Timing Results}

Figure \ref{xmmlc} shows summed X-ray and $\gamma$-ray light curves from 0.1 keV
-- 7.0 keV, and above 100 MeV, folded at the post-glitch ephemeris, given in
the second column of Table 2. These light curves can be compared with
Figure 4 of \cite{mj02}.

\begin{figure}
\center{\scalebox{0.65}{\plotone{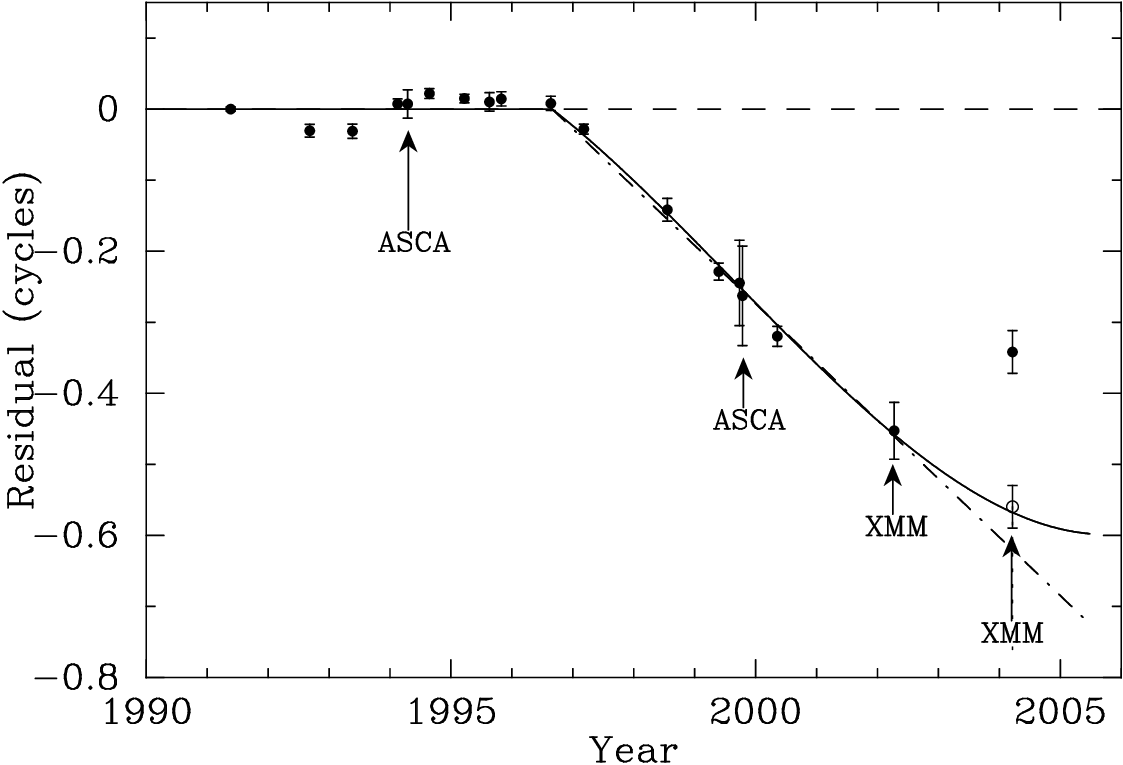}}}
\caption{Phase residuals of the EGRET timing observations of Geminga
  relative to the cubic ``1997 ephemeris'' of \cite{ma98} (pre-glitch
  ephemeris in Table \ref{tbl-2}).  The 14 unmarked measurements
  correspond to the numbered EGRET observations in Table \ref{tbl-1},
  some of which were grouped together, and the 1994 and 1999 \asca\
  GIS and 2002 and 2004 \xmm\ pn observations are indicated. The
  solid line represents the cubic ephemeris segments before and after
  the glitch. The dash-dot line denotes the previous post-glitch
  ephemeris given in \cite{mj02}, and the dashed line indicates
  phase zero of the pre-glitch ephemeris. The post-glitch ephemeris is
  also given in Table \ref{tbl-2}. The unfilled circle below the 2004
  \xmm\ point indicates the phase residual of the data that has
  been corrected for the possible 1-second error.\label{phaseplot}}
\end{figure}

\begin{figure}
\center{\scalebox{0.5}{\rotatebox{0}{\plotone{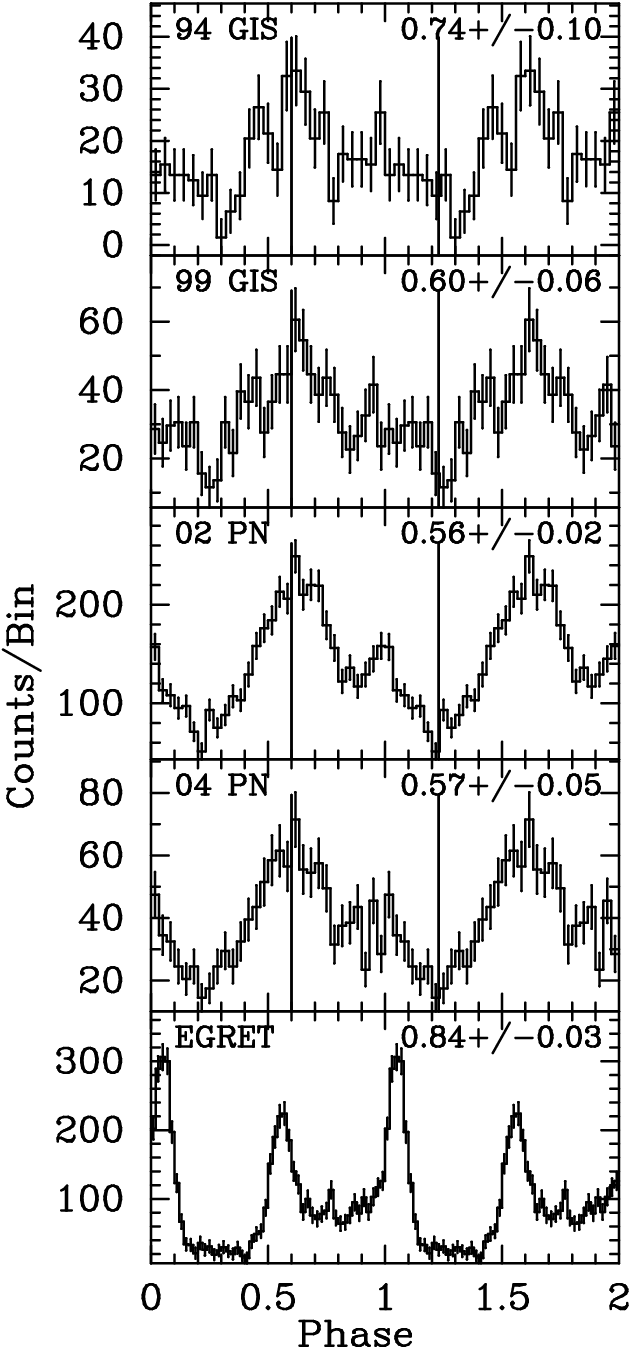}}}}
\caption{Light curves from the four X-ray observations (\asca\ GIS\ and
  \xmm\ pn) at 0.7--2.0 keV, and summed EGRET ({\it bottom panel})
  observations, folded at the applicable ephemeris given in Table
  2. The pulsed fraction for the light curves are shown in each
  panel. The vertical lines show how the peaks and troughs line up from
  light curve to light curve.\label{xraylc}} 
\end{figure}

\begin{figure}
\center{\scalebox{0.5}{\rotatebox{0}{\plotone{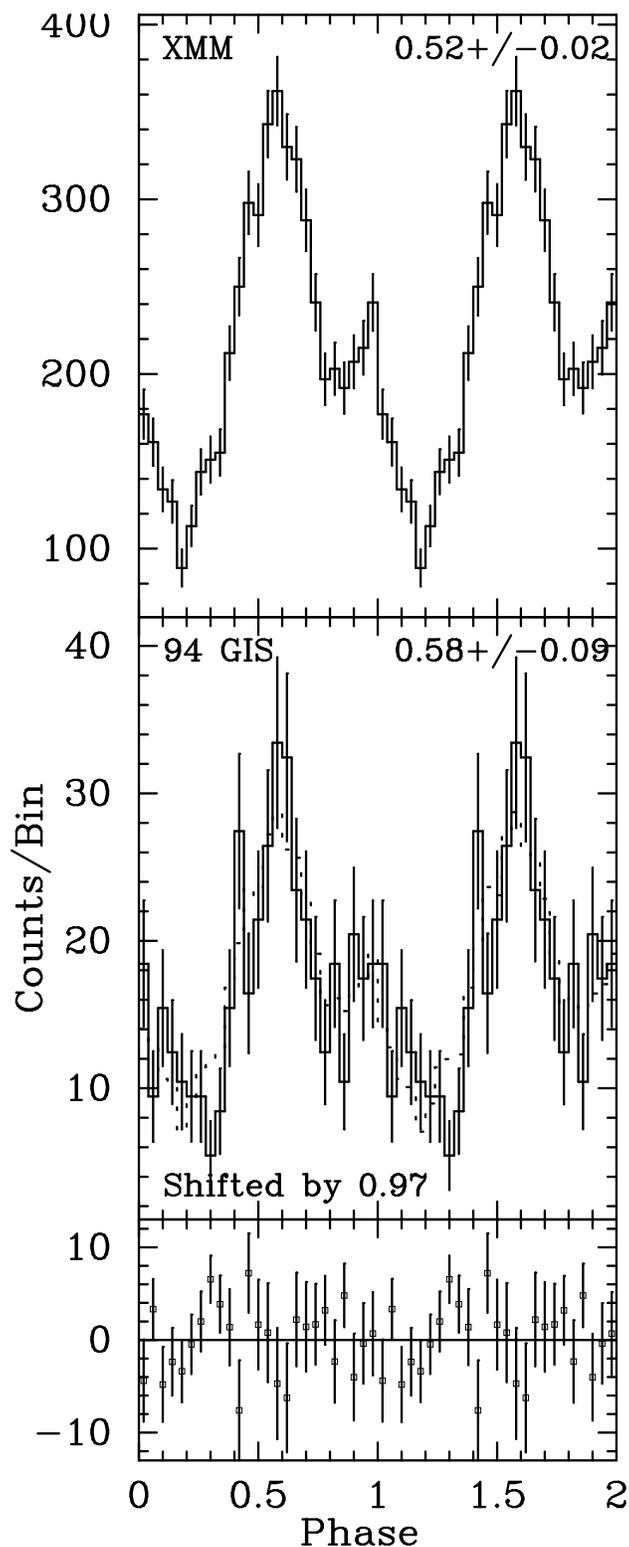}}}}
\caption{A comparison of the 1994 \asca\ GIS light curve ({\it middle panel})
  and XMM composite light curve ({\it top panel}) at 0.7--2.0 keV. The
  lower light curve is shifted
  in phase by the amount indicated in the lower left corner to best
  match the upper
  light curve. The dotted line shows the top light curve
  normalized to the bottom. The pulsed fractions are indicated in the
  upper right corner of the panels and the residuals are shown in the
  bottom panel. \label{complc}}
\end{figure}

\begin{figure}
\center{\scalebox{0.5}{\rotatebox{0}{\plotone{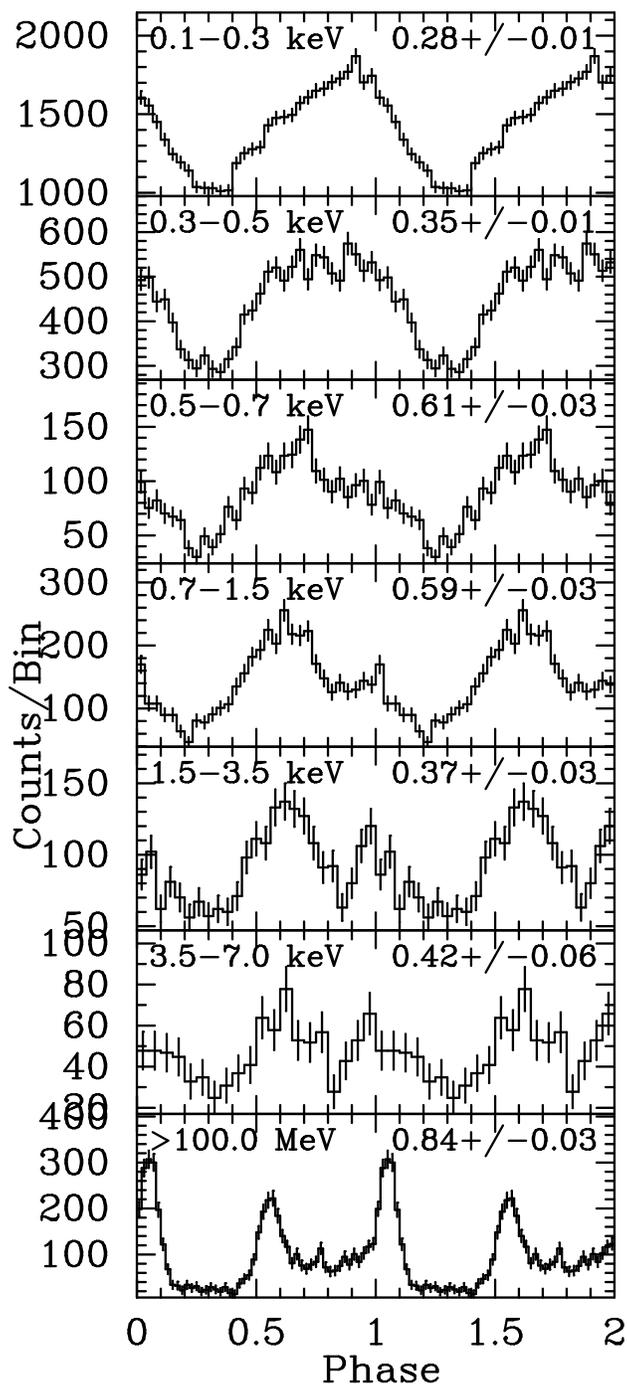}}}}
\caption{Light curves from the summed 2002 and 2004 \xmm\ pn ({\it top six 
panels}),
  and summed EGRET ({\it bottom panel}) observations, folded at the updated
  post-glitch ephemeris given in the second column of Table 2. The
  energy range
  and pulsed fraction for the light curves are shown in each panel.\label{xmmlc}}
\end{figure}

It can be seen from Figure \ref{phaseplot} that the Geminga ephemeris,
consisting of 2 cubic segments covering 1973--1996 and 1996--2004 given
in Table 2, is phase-connected and valid for the full specified epoch
range. Apart from instances of timing noise, which manifest themselves
as phase deviations from the solid line in Figure \ref{phaseplot}, the
only confirmed glitch since 1973 occurred in late 1996. In this paper,
the term ``pre-glitch'' refers to epochs before the 1996 glitch.

For the ephemeris search, the value of $\ddot f$ was assumed to be
zero, as that parameter would not have significantly affected the
fits, given the 1400 day timeline. Figure \ref{phaseplot} indicates
that when the $\ddot f$ parameter is zero, the ephemeris matches the
data well, and the phase residuals for the 2002 \xmm\ observation using
the post-glitch ephemeris are consistent with zero, also indicating
that the post-glitch ephemeris parameters, including the $\ddot f$
value, are valid up to at least 2002. Further \xmm\ observations will allow
for a determination of the ephemeris from X-ray data alone, using the
method described in Section 5. This will assure that the ephemeris is
consistent and phase-connected and will also determine whether a glitch occurred
between the 2002 and 2004 \xmm\ observations.

\section{Spectral Analysis}\label{specan}

\begin{figure}
\centering{
\scalebox{0.55}{\rotatebox{270}{\plotone{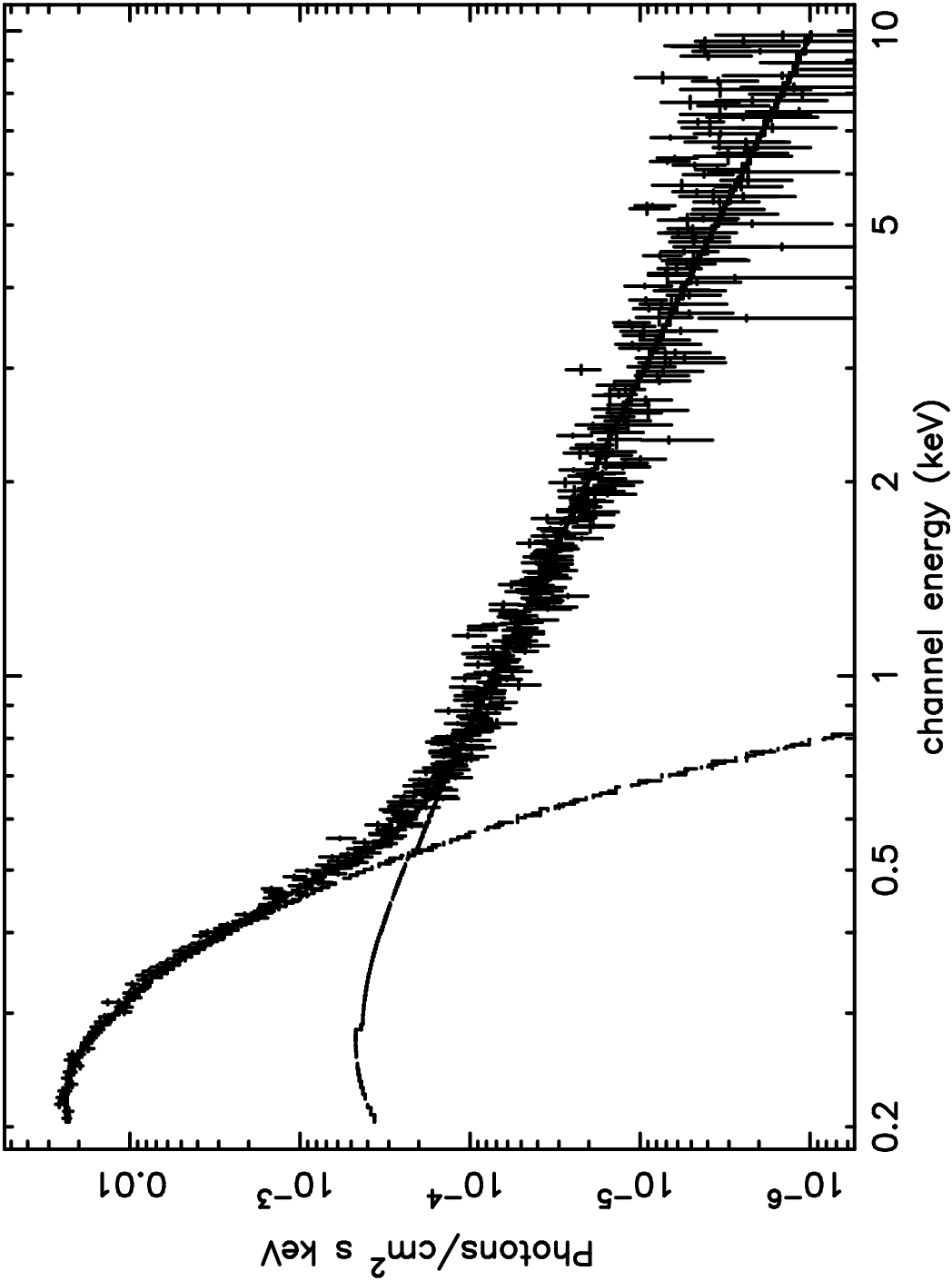}}}
\scalebox{0.55}{\rotatebox{270}{\plotone{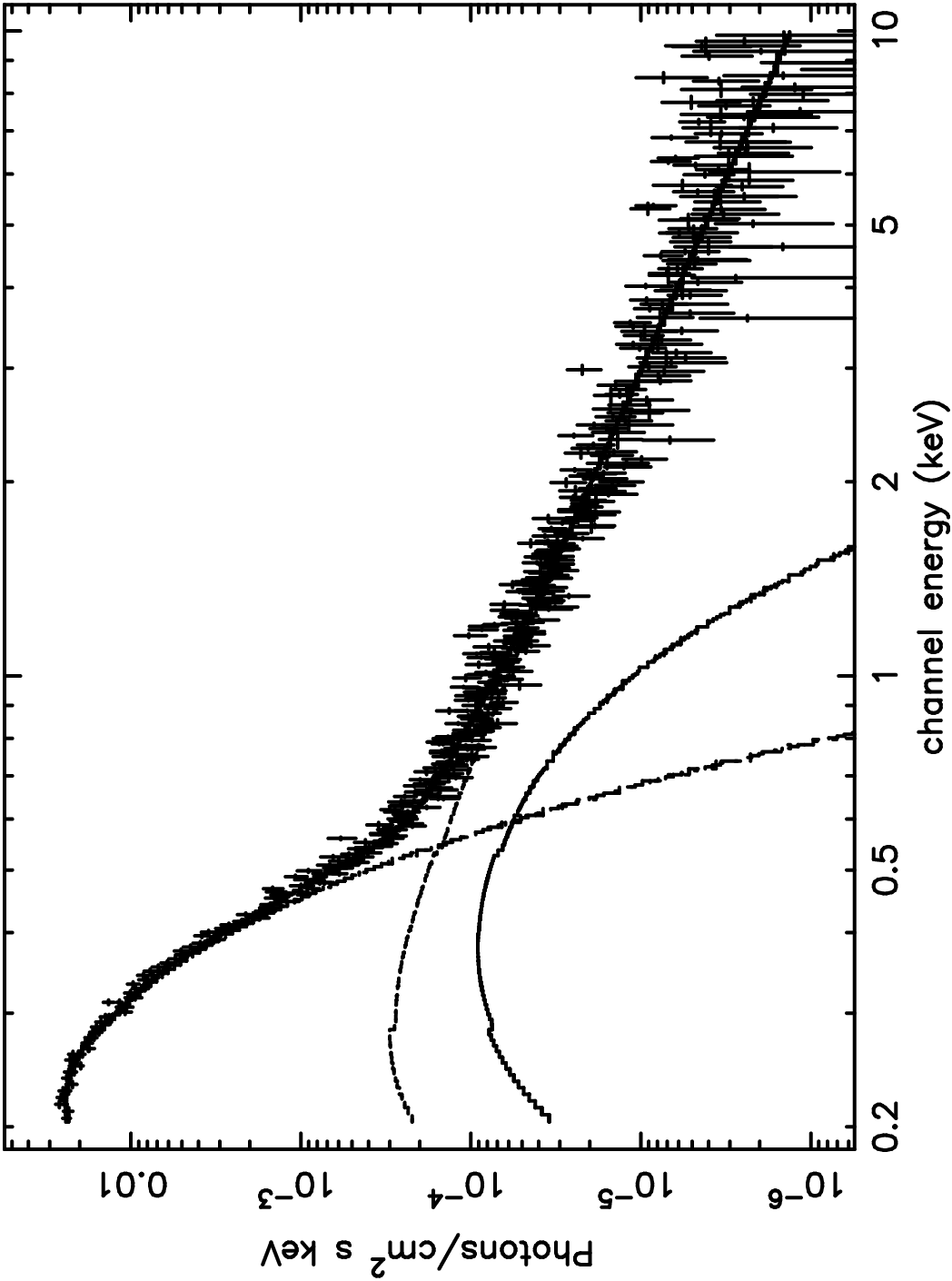}}}}
\caption{Fit of combined \xmm\ pn unfolded spectra of Geminga to {\it
  Top panel:} an absorbed power law
  plus single blackbody (Model A in text); {\it
  Bottom panel:} an absorbed power law plus warm and hot blackbody
  components (Model B in text). The power law and blackbody components are
  shown in addition to the data and total model. 
The fitted parameters are given in Table 3. \label{speca}}
\end{figure}

The combined 2002 and 2004 \xmm\ EPIC-pn full spectra, extracted as described in
Section \ref{dataprep}, were fitted between energies of 0.2
and 10.0 keV, to an absorbed power law and one or two blackbody
components. The spectra showing the power law and blackbody components
are given in Figure \ref{speca} and the fitted
parameters are given in Table 3.  

The values of the power law index are consistent with those from
previous \asca\ measurements \citep{mj02} and the previously published
value from the 2002 \xmm\ data \citep{ca04}, and between the 2002 \xmm\
measurement alone and the combined 2002 and 2004 fit, but the addition
of the hot blackbody component decreases the value of this
parameter. The fitted temperature of the warm blackbody component is
consistent with the \ro\ and \asca\ values \citep{hw97}, and both warm
and hot blackbody temperatures are consistent with the previously
published values from the 2002 observation \citep{ca04}. 

The poor response of the detector at low energies results in a large
error bar on the column density and the fit of that parameter is
strongly coupled to the parameters of the warm blackbody
component. However, the value is consistent among fits of one and  two
blackbody components to the 2002 observation alone and the combined
2002 and 2004 data, and larger than (but within uncertainty of) those
that arose from fits performed on \ro\ and combined \ro\ and \asca\
data \citep{hw97,hr93}. It was noted at the time of those measurements
that the \ro\ PSPC entrance window was becoming progressively thinner,
allowing more low energy photons to pass through and leading to a
smaller and smaller measured column density. The true column density
is therefore most likely greater than the 1.07$\times10^{20}$
cm$^{-2}$ value from \ro\ data or the 1.38$\times10^{20}$ cm$^{-2}$
value from combined \ro\ and \asca\ data given in \cite{hw97}.

As is shown in Table 3, combining the 2004 spectrum with that of the
2002 observation has the effect of reducing the error bars on the
parameters. Although the reduced $\chi^2$ value increases slightly, the
fit to the combined data results in parameters very consistent with
those from the 2002 observation alone.

\begin{figure}
\centering{\scalebox{0.5}{\rotatebox{270}{\plotone{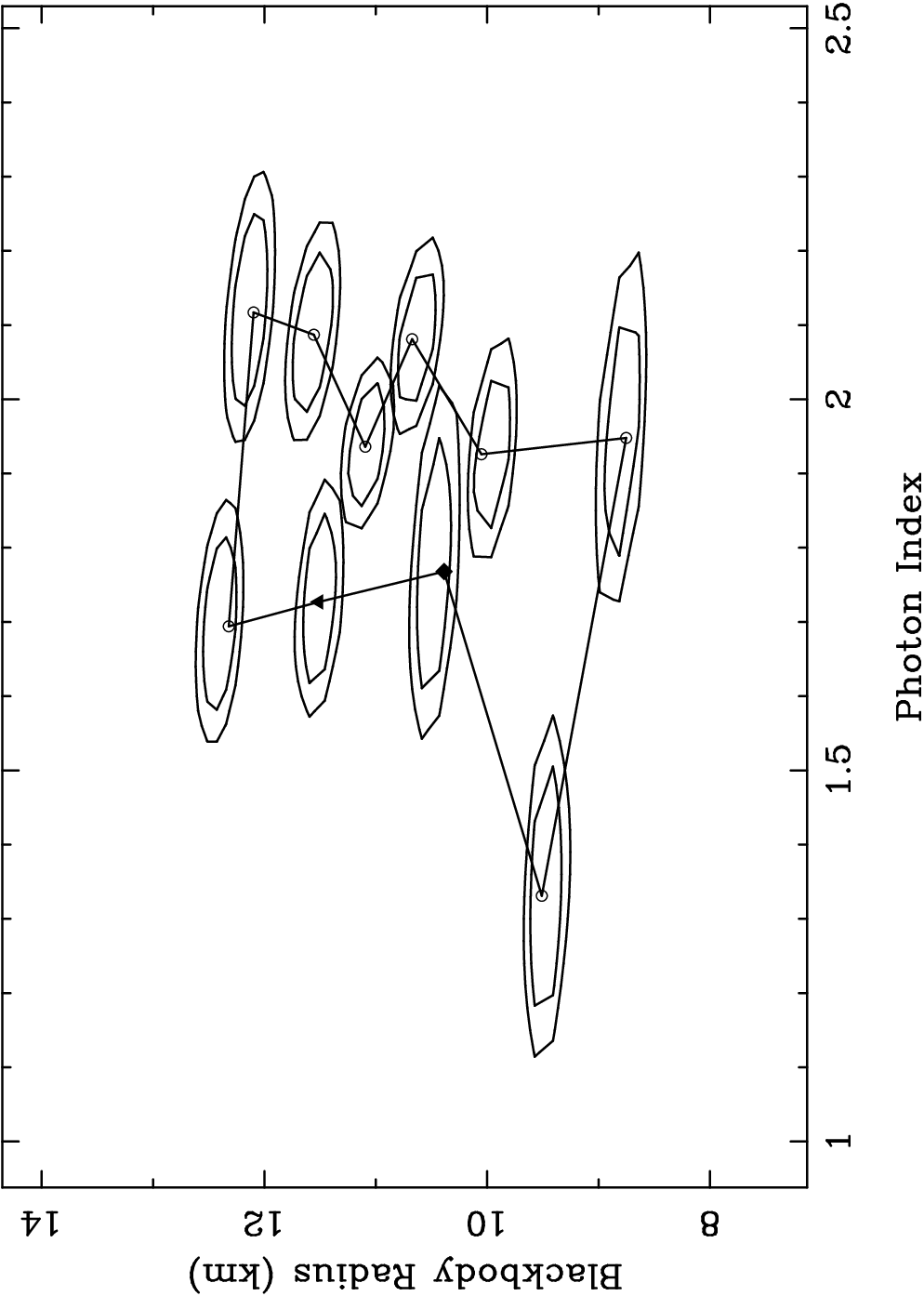}}}
\scalebox{0.5}{\rotatebox{270}{\plotone{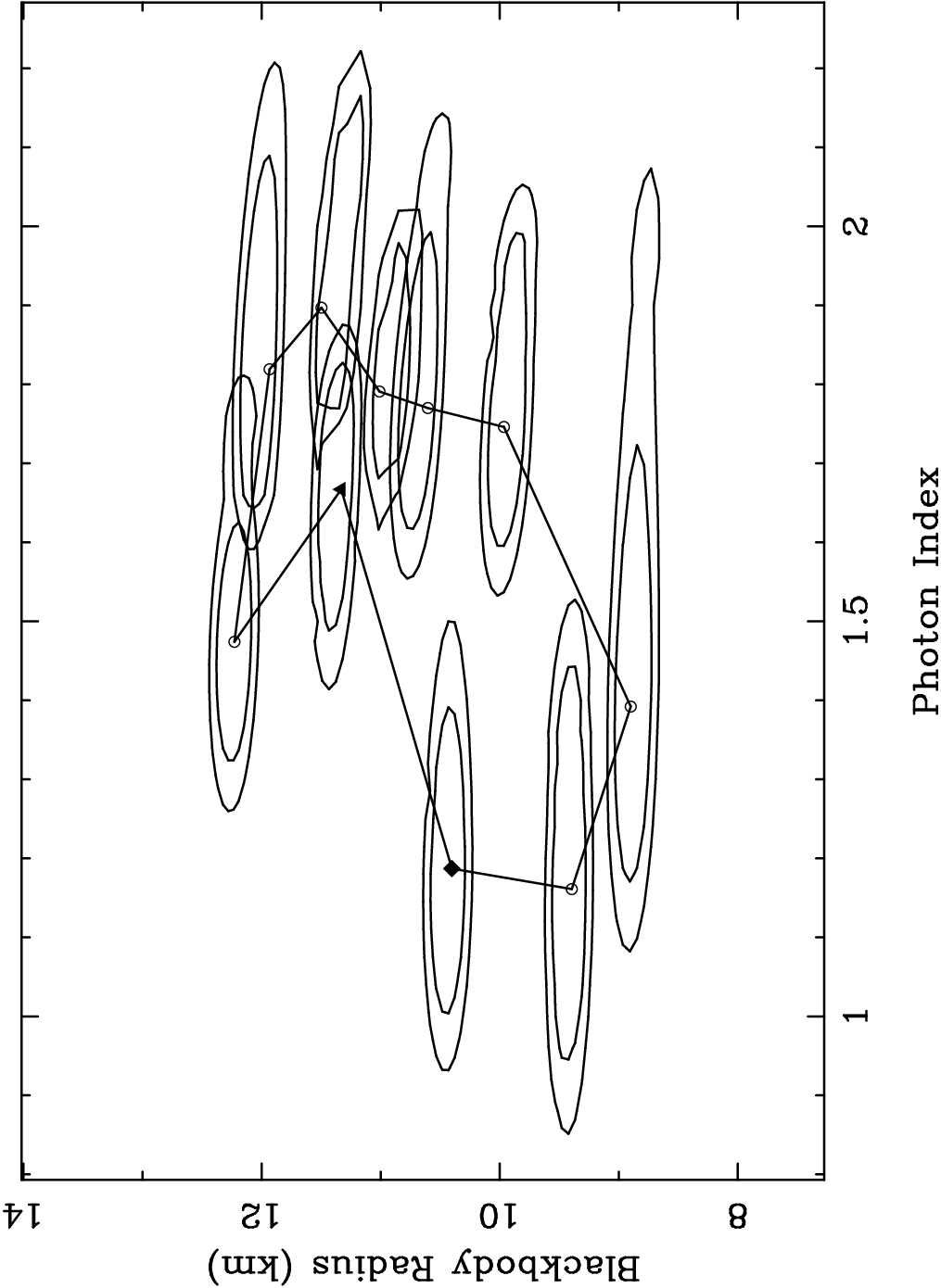}}}}
\caption{Contours of the photon index vs. normalization of the
  primary (warm) blackbody component. The contours are at 2 and 3
  $\sigma$ for two interesting parameters. The point 
  corresponding to phase 0.0--0.1 is shown as a triangle and the
  diamond indicates phase 0.1--0.2, with the adjacent phases
  connected. The top panel shows the contours for the single blackbody
  model (Model A) and the bottom panel shows the contours for the
  two blackbody model (Model B). \label{phovsbba}} 
\end{figure}

To fit the phase-resolved spectra using the one and two blackbody plus
power law models, the column density was frozen at its fitted value
from the fits to the entire spectrum. Data between 0.2 and 8.0 keV
were used for these fits, as the statistics at greater energies are
not sufficient to make reliable fits. For the first fits to the
phase-resolved spectra, all parameters other than the column density
were allowed to vary. Whereas \cite{ca04} noted that the photon index
did not show significant variation over the rotation cycle, we observe
significant variability of the photon index for fits of both one and
two blackbody components, as shown in the 2 and 3 $\sigma$ contours in
Figure \ref{phovsbba}. Whereas it appears that the lower panel of Figure 7 shows that as few as two points differ from the mean value as well as any possible chosen photon index value by $3\sigma$ or more, this would occur with less than 0.1\% probability if the photon index were in fact constant. Another fact supporting
the reality of the variability of the spectral index is that the values trace
out a loop in Figure 7 (both top and bottom panels).   That is, not only do the values differ by a few
$\sigma$, but they differ in a non-random order.  This behavior must enhance
the statistical significance of the variation, although it is not as easy to
quantify.

\begin{figure}
     \centering
\scalebox{0.80}{\plottwo{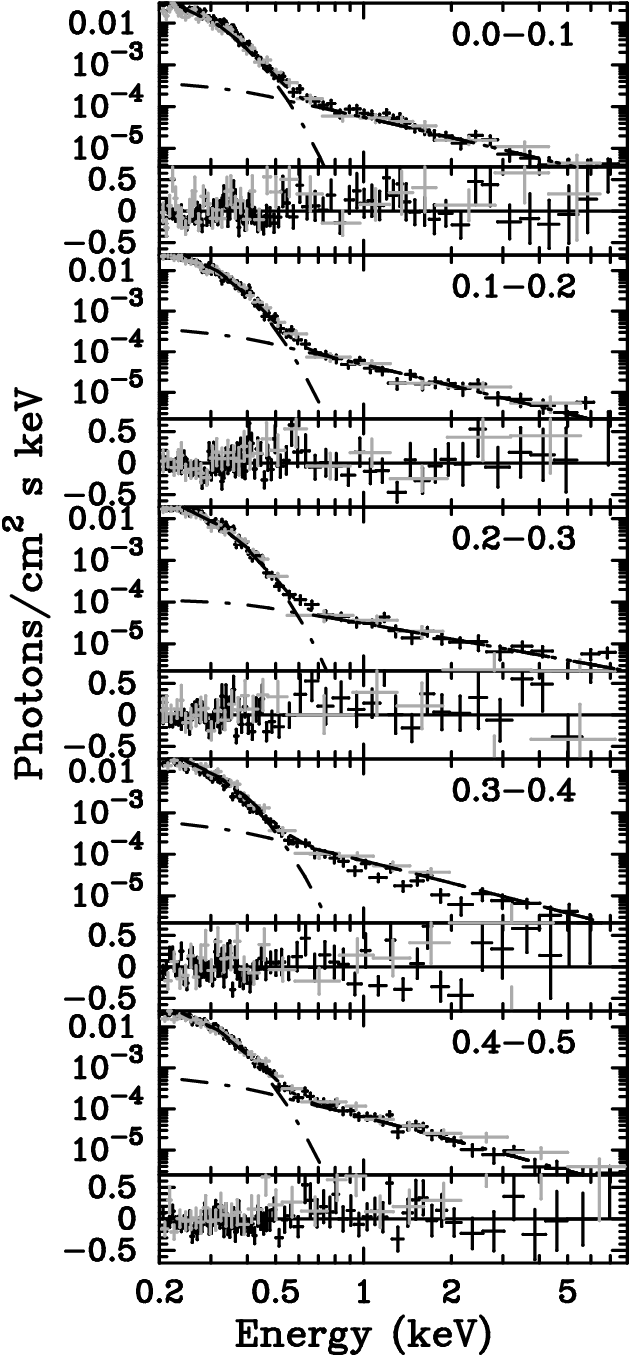}{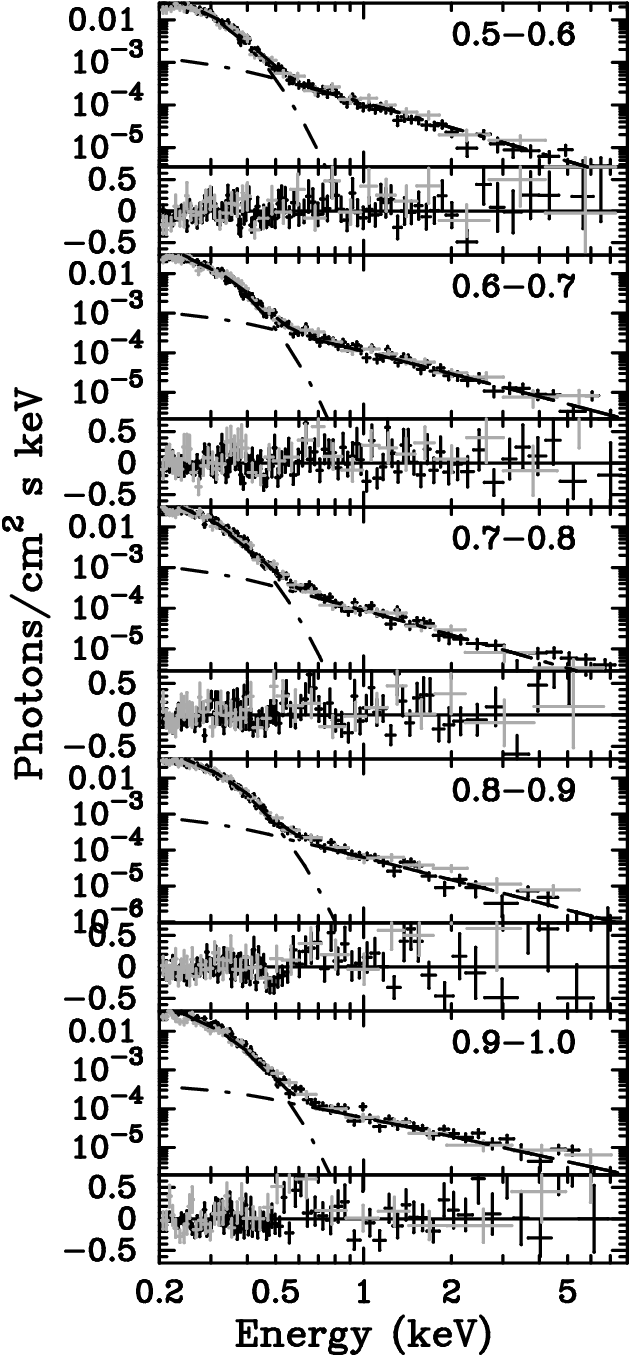}} 
     \caption{Fits of \xmm\ pn phase-resolved spectra to an absorbed
  power law plus single blackbody (Model A), for phase ranges
  indicated in the upper corner of each spectrum. The black points
  correspond to the 2002 observation and the light grey points
  correspond to the  2004 observation. The fitted parameters are given
  in Table 4. The power law and blackbody components are shown with
  the spectra, as well as the total model. The  fractional differences
  between the data and the model are shown below the spectra. \label{phspeca} }
\end{figure}

\begin{figure}
     \centering
\scalebox{0.80}{\plottwo{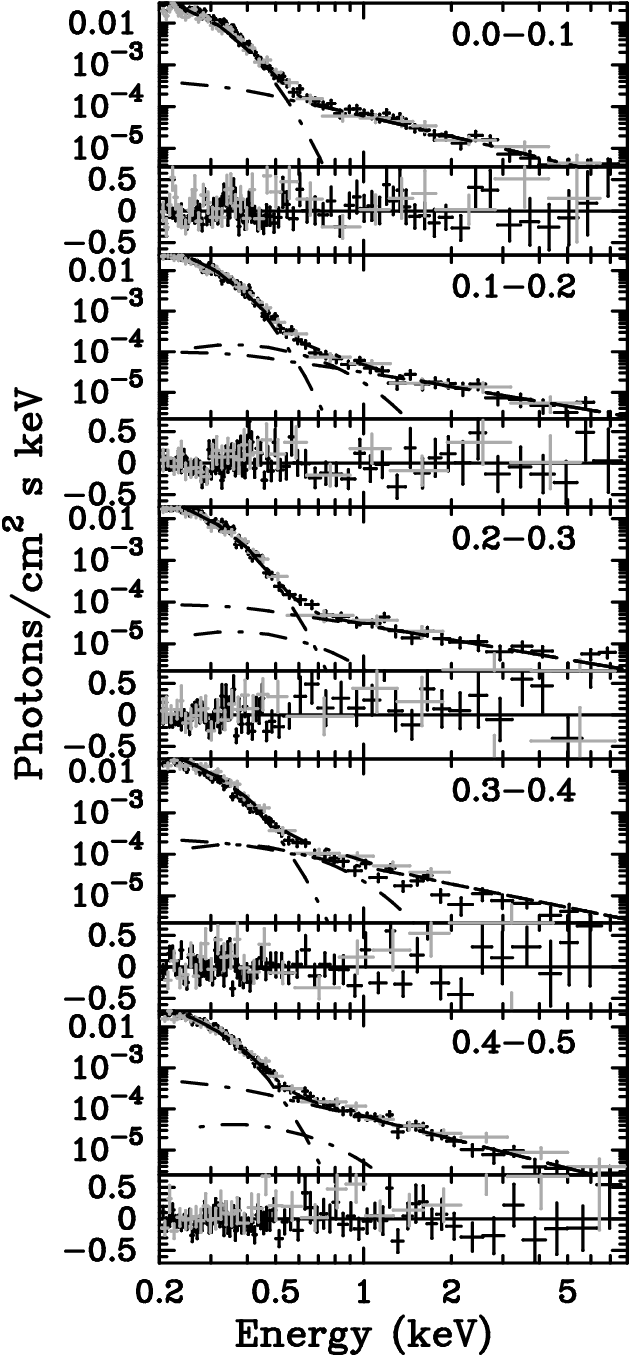}{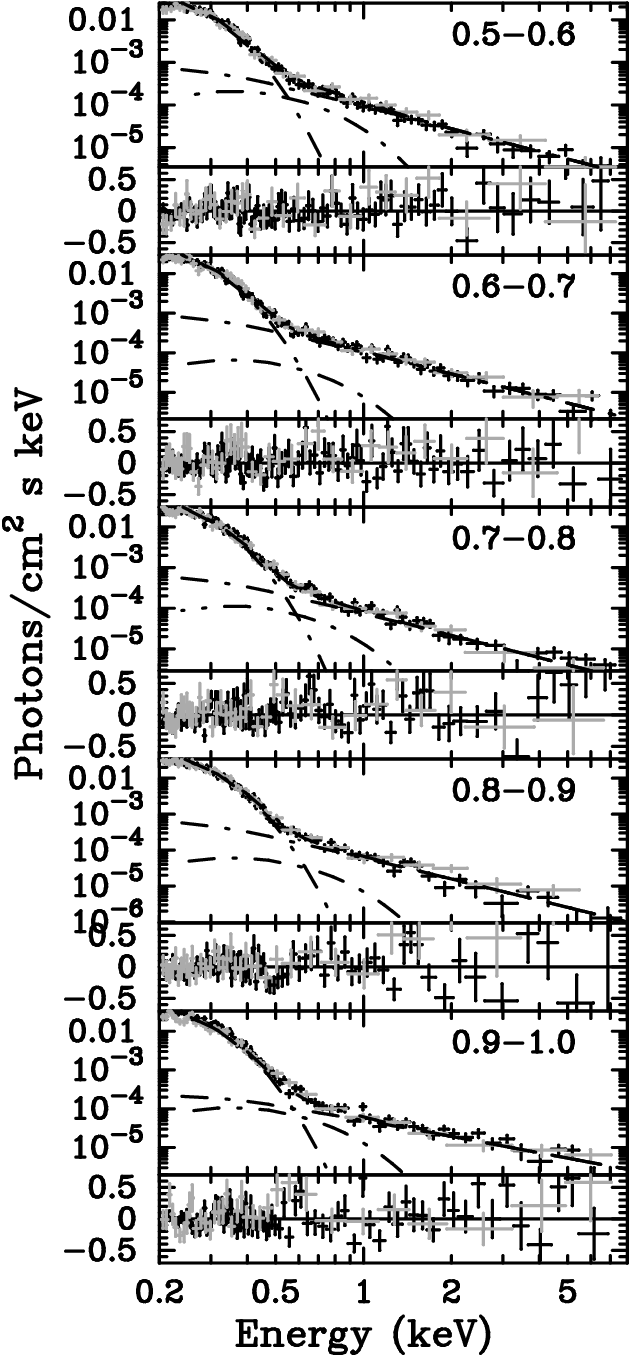}} 
\caption{Same as Figure \ref{phspeca}, for an absorbed power law
  plus warm and hot blackbody components (Model B).}
     \label{phspecb}
\end{figure}

We do confirm the result
of \cite{ca04} that the blackbody temperatures are constant within
uncertainty, as would be expected for rotating hot or warm spots, and
thereafter these temperatures were frozen for the fits. The absorbed
power law with a variable photon index plus one or two blackbody
components with fixed temperatures are hereafter referred to as Models
A \& B. 
The resulting fitted parameters for both models (Models A \& B) are
given in Table 4, and the fitted phase resolved spectra and their
ratio to the model are shown in Figures \ref{phspeca} and
\ref{phspecb}. The values of the  reduced $\chi^2$ are not
significantly decreased with the addition of the second blackbody
component (Model B) and in some cases it actually increases (though
the value of $\chi^2$ does not increase), and most of the
normalization values for the hot  blackbody component are consistent
with zero, considering the rather large error bars, so it is not clear
that the hot blackbody component is present in the spectra at all,
except in one or two phase bins where the value of the emitting radius
exceeds its error bar. 

\begin{figure}
     \centering
\scalebox{0.80}{\plottwo{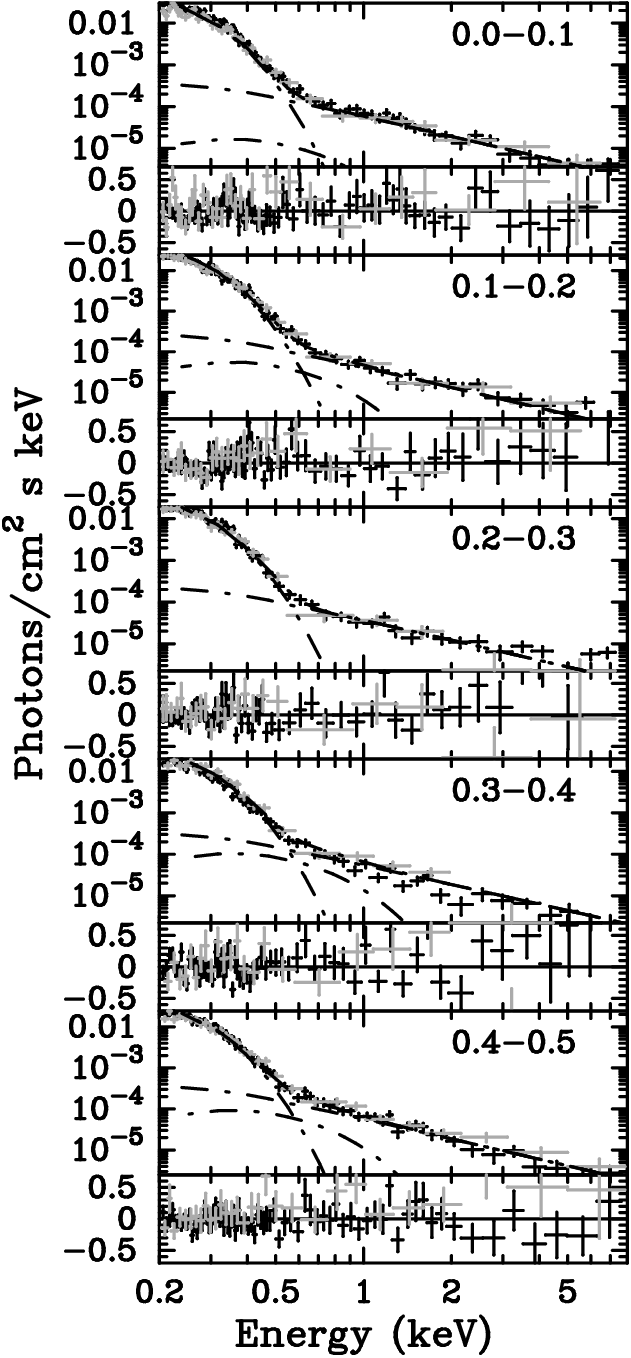}{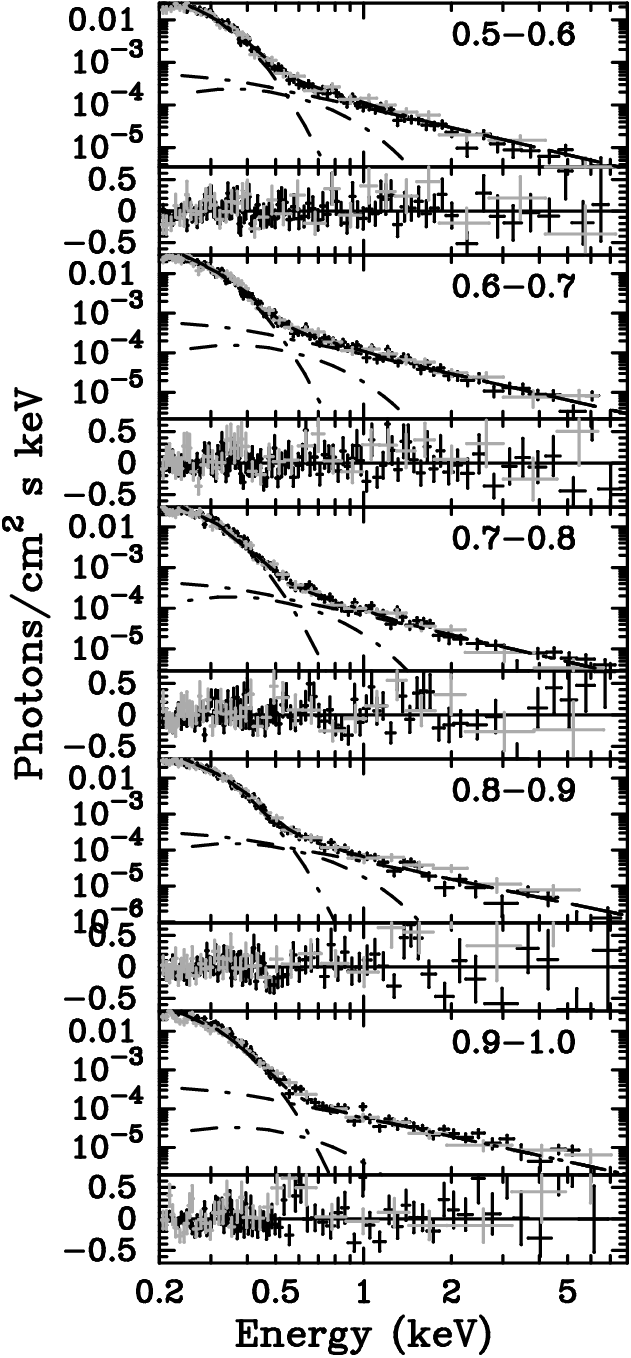}} 
\caption{Same as Figure \ref{phspecb}, with a fixed photon index (Model C).}
     \label{phspecc}
\end{figure}

As a test of the consistency between the results given here and the
previously published results \citep{ca04}, the method used in that
paper was employed here. For the fits to the phase resolved spectra,
the two blackbody plus power law model was employed, with the power
law index fixed at the value obtained for the full spectrum, and
everything else the same as in Model B, i.e. the temperatures and
column density frozen as well, leaving only the normalizations of the
three components to vary. This model is hereafter referred to as Model
C. The results are given in Table 4, and the parameters are similar to
those in \cite{ca04}, with the the hot blackbody component more
significant than when the power law index was allowed to vary. The
fitted phase resolved spectra and their ratio to the model (Model C)
are shown in Figure \ref{phspecc}. 

From Figures \ref{phspeca}, \ref{phspecb}, and \ref{phspecc}, it is
clear that Models A, B, and C fit the data approximately equally well
for each phase-resolved spectrum, especially at energies below 5 keV,
and the unfitted features in some of the phase bins are apparent  for
all of the models. For example, for the 0.8--0.9 phase bin, there is a
dip in the spectrum compared to the model at 0.5 keV, and for the
0.2--0.3 and 0.3--0.4 phase bins, there is a narrow hump between 0.3 and
0.4 keV. Since these features are apparent in both the 2002 and 2004
data sets, these are more likely actual spectral features that are
not well-fit by the model being used, than remnants of poor
statistics in the phase-resolved spectra. Clearly there is an as yet
unexplained process that emits in X-rays, but it is not described by
the models employed here, and it could have its origin either on the 
neutron
star surface or in the magnetosphere.

\begin{figure}
\center{
\scalebox{0.5}{\rotatebox{0}{\plotone{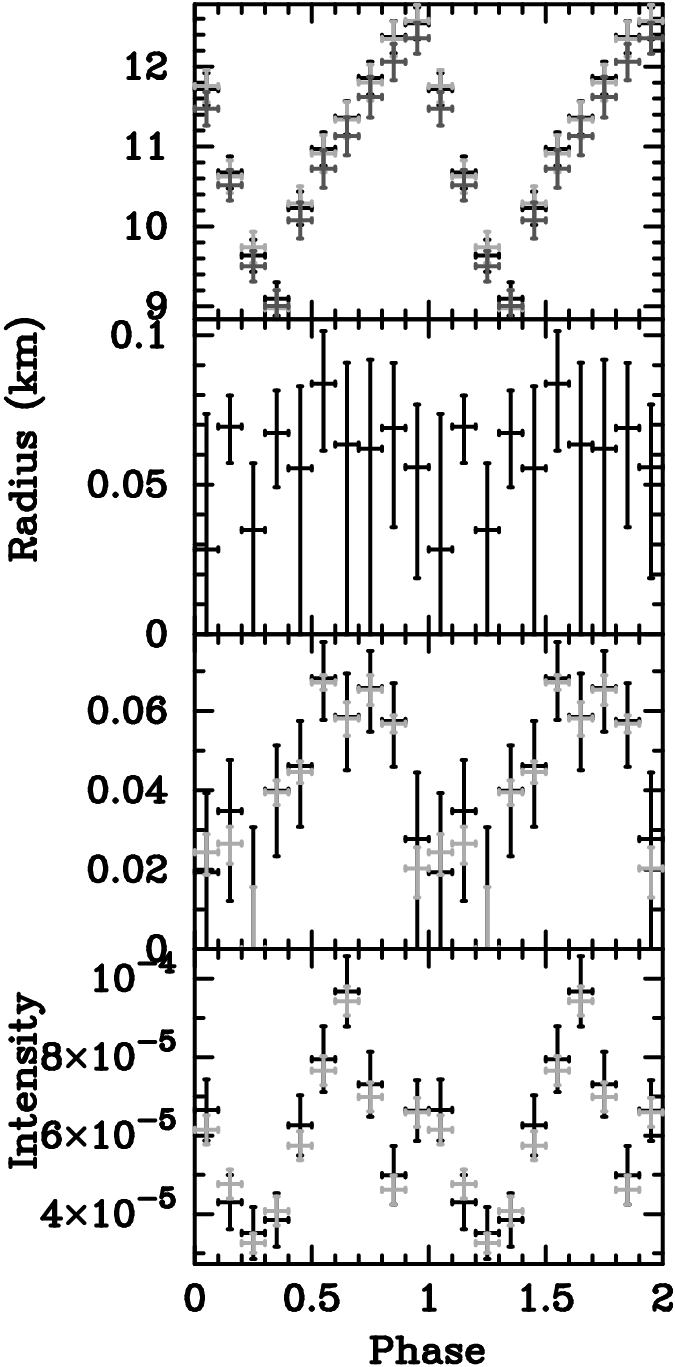}}}
}
\caption{Light curves for the modulation of individual spectral
  components. The top
  panel shows the modulation of the warm blackbody emitting radius
  for Models A ({\it light grey}), B ({\it dark grey}), and C ({\it black}). The second
  panel shows 
  the modulation of the hot blackbody emitting radius for Model B. The
  third panel shows the hot blackbody emitting radius for
  Model C ({\it black}) and fake data ({\it light grey}). The
  the bottom panel shows the modulation of the power law for Model C
  ({\it black}) and the fake data ({\it light grey}).\label{bblcab} }
\end{figure}

The variation of
the emitting radii of the blackbody components and power law intensity for the models are
shown in Figure \ref{bblcab}. The warm
blackbody component is prominent at the lowest energies, and the power
law component dominates at higher energies. When the power law index
is fixed, the hot blackbody component modulates approximately with the
power law component, and not with the warm blackbody component as
would be expected from a model where the neutron star surface is
heated from the interior along the field lines near the poles.

A possible explanation for the behavior of the spectral parameters in
the fits, specifically the variation of the hot blackbody component in
Model C, was tested by creating a fake spectrum for each of the 10
phase bins with the parameters given in the first column of Table 4
(the power law plus single blackbody, Model A). Using these fake
spectra as input, fits were made using a model containing a power law
and two blackbody components, with temperatures and photon index fixed
as they were for Model C. The
resulting parameters are very similar to those obtained from fitting
the real spectra to Model C. It is quite
evident from the similarity between the parameters that the varying
photon index in the true spectra produces an excess of residual
counts in fits where the photon index is held fixed, and these excess
counts can be fit by a second blackbody component. That does not mean
the hot blackbody component is present in the spectra but it could be
the result of erroneously freezing the photon index for the fits where
it is clear from Figure \ref{phovsbba} that it in
fact varies at a $>99.9\%$ confidence level.

\section{Results of Phase-resolved Spectroscopy}

While there is some evidence for a variable hot blackbody component in the
\xmm\ Geminga phase resolved spectra when the photon index is held
fixed, we find that the strength of the variation is not nearly as
clear-cut as has been previously stated \citep{ca04}. The residuals
from a single blackbody plus power law fit could arise from the
magnetospheric emission not being accurately described by a power
law. This argument is strengthened by the fact that the hot blackbody
normalization for Model C closely matches the strength of the power
law component in the light curves. The
\xmm\ data do not prove conclusively that there is a rotating hot spot
on the neutron star surface, and it is more likely that
the emission from the hot spot would have a low pulsed fraction, than
that it completely disappears for 1/10 of the rotation. The two-blackbody 
fits to the fake spectra, which were made to conform to the
fitted parameters of the single  blackbody fits, result in parameters
that closely match those from similar fits to the real data, and the
apparent disappearance of the hot blackbody component for 1/10 of the
phase was due to the fact that the correction to the fixed photon
index model provided by the addition of the second blackbody was not
necessary for that particular dataset.

The method used in this and previous papers, i.e. to fit the entire
spectrum to a single power law and one or two blackbody components,
and then to use some of the parameters for subsequent fits to the
phase-resolved data, does not take into account the fact that the sum
of power laws is not itself a power law. While it is unlikely that the
magnetospheric emission is a power law, even over the limited 0.2--8 keV
energy range for a fraction of the rotation, the full spectrum, which
is a sum of the emission for the full period, is even less likely to
be a power law. Unfortunately, the exposure times of the
observations used for this paper were not sufficient to accumulate
spectra that will answer the question of the exact spectral shape of
the phase-resolved spectra, but the data provide a much better picture
of the variation of the spectrum over the rotation of the pulsar, and
it can safely be concluded that both the magnetospheric and surface
emission vary significantly. 

It is not clear whether the non-thermal component of the X-ray
spectrum of Geminga can be extrapolated to the ultraviolet and optical
range, and it has already been established in previous work that the
spectrum does not extrapolate to $\gamma$-ray energies
\citep{mj02}. The power law index in the X-ray regime is less
uncertain with the new \xmm\ observations, but it strongly depends on
other parameters such as the column density and the strength and
temperature of the warm and hot (if present) blackbody components. As
previously stated, the non-thermal component of the spectrum is
parameterized by a power law, but it is not certain that the spectrum
is a power law or that it would continue with the same slope at all
energies. 

\acknowledgments

{\it XMM-Newton} is an ESA science mission with instruments and
contributions directly funded by ESA Member States and NASA.
This research was supported by NASA grant NNG04GH83G.

\clearpage

\clearpage

\clearpage

\clearpage 

\clearpage 

\clearpage 

\clearpage 

\clearpage

\clearpage 

\clearpage 

\clearpage

\clearpage

\begin{deluxetable}{llrr}
\tabletypesize{\scriptsize}
\tablecaption{Log of Observations \label{tbl-1}}
\tablewidth{0pt}
\tablehead{
\colhead{Instrument} & \colhead{Dates} & \colhead{Exposure time (ks)} &
\colhead{Count rate (s$^{-1}$)} }
\startdata
EGRET (1) & 1991 Apr 22--May 7 & 1209.6 & 1.8$\times 10^{-3}$ \\
EGRET (1) & 1991 May 16--30 & 1209.6 & 1.6$\times 10^{-3}$ \\
EGRET (1) & 1991 Jun 8--15 & 604.8 & 1.5$\times 10^{-3}$ \\
EGRET (2) & 1992 Jun 11--25 & 1209.6 & 2.2$\times 10^{-4}$ \\
EGRET (2) & 1992 Aug 11--20 & 777.6 & 1.9$\times 10^{-4}$ \\
EGRET (2) & 1992 Sep 1--17 & 1382.4 & 1.7$\times 10^{-4}$ \\
EGRET (2) & 1992 Oct 8--15 & 604.8 & 1.4$\times 10^{-4}$ \\
EGRET (2) & 1992 Nov 3--17 & 1209.6 & 1.0$\times 10^{-4}$ \\
EGRET (3) & 1993 Mar 23--29 & 604.8 & 3.2$\times 10^{-4}$ \\
EGRET (3) & 1993 May 13--24 & 950.4 & 4.1$\times 10^{-4}$ \\
EGRET (4) & 1993 Dec 1--13 & 1036.8 & 3.5$\times 10^{-4}$ \\
EGRET (4) & 1994 Feb 8--17 & 777.6 & 7.7$\times 10^{-4}$ \\
EGRET (5) & 1994 Aug 9--29 & 1814.4 & 3.7$\times 10^{-4}$ \\
EGRET (6) & 1995 Feb 28--Mar 21 & 1814.4 & 4.0$\times 10^{-4}$ \\
EGRET (6) & 1995 Apr 4--11 & 604.8 & 2.5$\times 10^{-4}$ \\
EGRET (6) & 1995 May 9--Jun 6 & 2419.2 & 1.8$\times 10^{-4}$ \\
EGRET (7) & 1995 Aug 8--22 & 1209.6 & 2.4$\times 10^{-4}$ \\
EGRET (8) & 1995 Oct 17--31 & 1209.6 & 2.7$\times 10^{-4}$ \\
EGRET (9) & 1996 Jul 30--Aug 27 & 2419.2 & 3.4$\times 10^{-4}$ \\
EGRET (10) & 1997 Feb 18--Mar 18 & 2419.2 & 1.5$\times 10^{-4}$ \\
EGRET (11) & 1998 Jul 7--21 & 1209.6 & 2.0$\times 10^{-4}$ \\
EGRET (12) & 1999 May 11--25 & 1209.6 & 2.1$\times 10^{-4}$ \\
EGRET (13) & 1999 Sep 14--28 & 1209.6 & 4.7$\times 10^{-5}$ \\
EGRET (14) & 2000 Apr 25--May 9 & 1209.6 & 2.6$\times 10^{-4}$ \\
ASCA SIS & 1994 Mar 28--31 & 49.2 & 1.7$\times 10^{-2}$ \\
ASCA GIS & 1994 Mar 28--31 & 75.3 & 1.2$\times 10^{-2}$ \\
ASCA SIS & 1999 Oct 5--11 & 194.0 & 1.4$\times 10^{-2}$ \\
ASCA GIS & 1999 Oct 5--11 & 207.8 & 1.0$\times 10^{-2}$ \\
XMM pn & 2002 Apr 4--5 & 71.4 & 0.67 \\
XMM MOS 1 & 2002 Apr 4--5 & 101.9 & 0.12 \\
XMM MOS 2 & 2002 Apr 4--5 & 101.9 & 0.12 \\
XMM pn & 2004 Mar 13 & 18.0 &  0.71 \\
XMM MOS 1 & 2004 Mar 13 & 26.0 & 0.15 \\
XMM MOS 2 & 2004 Mar 13 & 26.0 & 0.16 \\
\enddata

\end{deluxetable}

\clearpage 

\begin{deluxetable}{lll}
\tabletypesize{\scriptsize}
\tablecaption{The Geminga EGRET Ephemeris \label{tbl-2}}
\tablewidth{0pt}
\tablehead{\colhead{Parameter} & \colhead{Pre-glitch\tablenotemark{a}} & \colhead{Post-glitch}}
\startdata
Epoch of ephemeris $T_0$ (MJD)\tablenotemark{b}  & 46599.5 & 50497.718748124877\\
Range of valid dates (MJD)  &  $41725-50320$ & $50320-53078$\tablenotemark{c} \\
Frequency $f$ (Hz) & $4.217705363081(13)$ & $4.217639623538(35)$\\
Frequency derivative $\dot f$ (Hz s$^{-1}$) & $-1.9521712(12) \times 10^{-13}$ & $-1.9515522(81) \times 10^{-13}$ \\
Frequency second derivative $\ddot f$ (Hz s$^{-2}$) & $1.49(3) \times
10^{-25}$ & 0 \\
   \tableline
   \multicolumn{1}{c}{Parameter\tablenotemark{d}} &\multicolumn{1}{c}{Value} & \\
   \tableline
Epoch of position (MJD)	& 49793.5 	& \\
R.A. (J2000) &	$6^{\rm h}33^{\rm m}54.\!^{\rm s}153$ & \\
Decl. (J2000) &	$+17^{\circ}46^{\prime}12.\!^{\prime\prime}91$ &  \\
R.A. proper motion $\mu_{\alpha}$ (mas yr$^{-1}$)	&  138 & \\
Decl. proper motion $\mu_{\delta}$ (mas yr$^{-1}$)	&  97  & \\
\enddata
\tablenotetext{a}{From \cite{ma98}.}
\tablenotetext{b}{Epoch of phase zero in all light curves}
\tablenotetext{c}{The post-glitch ephemeris is provisional after the 2002 observation (MJD 52369), depending upon whether a glitch occured between then and the 2004 observation.}
\tablenotetext{d}{Position and proper motion from \cite{ca98}.}
\tablecomments{Digits in parentheses following a parameter value indicate $\sim$95\% confidence uncertainties 
in the last digits of the parameter.}
\end{deluxetable}

\clearpage

\begin{deluxetable}{lcccc}
\tabletypesize{\scriptsize} 
\tablecaption{Fits to XMM pn Spectra \label{tbl-3}}
\tablewidth{0pt} 
\tablehead{
\colhead{Parameter} & \multicolumn{2}{c}{1 Blackbody} &
\multicolumn{2}{c}{2 Blackbody}\\
\colhead{} & \colhead{2002 Observation} & \colhead{2002+2004 Combined}
& \colhead{2002 Observation} & \colhead{2002+2004 Combined}}  
\startdata 
$n_{\rm H}$ ($10^{20}$cm$^{-2}$) & 1.75 $\pm$ 0.63 & 1.75 $\pm$ 0.56 &1.76 $\pm$ 1.11&1.76 $\pm$ 0.95\\
$\Gamma$ & 1.901 $\pm$ 0.038& 1.895 $\pm$ 0.034& 1.697 $\pm$ 0.067&
1.684 $\pm$ 0.060\\
PL Normalization\tablenotemark{a} &7.44 $\pm$ 0.39  & 7.56 $\pm$ 0.35 &  6.22 $\pm$ 0.85&  6.30 $\pm$ 0.76\\
$T_{\rm w}$ (10$^5$ K) & 4.804 $\pm$ 0.015 & 4.818 $\pm$ 0.013 &4.773 $\pm$ 0.022 &4.800 $\pm$ 0.020\\
$R_{\rm w}$ (km) & 11.0424 $\pm$ 1.46 & 11.0297 $\pm$ 1.11 & 11.35 $\pm$ 0.87 &
11.17 $\pm$ 1.09\\
$T_{\rm h}$ (10$^5$ K) & \nodata & \nodata &17.0 $\pm$ 2.6 &17.1 $\pm$ 2.3\\
$R_{\rm h}$ (m) & \nodata & \nodata & 62. $\pm$ 39. & 62. $\pm$ 34.\\
Reduced $\chi^2$ & 1.189 (316 dof) & 1.208 (468 dof) & 1.131 (314 dof)
& 1.165 (466 dof)\\
\enddata 
\tablenotetext{a}{10 $^{-5}$ photons keV$^{-1}$cm$^{-2}$s$^{-1}$ at 1 keV}

\end{deluxetable}

\clearpage

\begin{deluxetable}{lcccccccl}
\tabletypesize{\scriptsize} 
\tablecaption{Fits to XMM pn Phase Resolved Spectra.\label{tbl-4}}
\tablewidth{0pt} 
\tablehead{
\colhead{Phase} & \colhead{Model} & \colhead{$\Gamma$} &
\colhead{PL Normalization at 1 keV} &
\colhead{$T_{\rm w}$} & \colhead{$R_{\rm w}$} &\colhead{$T_{\rm h}$} &
\colhead{$R_{\rm h}$} & \colhead{Reduced $\chi^2$ (dof)}
\\
& & & (10 $^{-5}$ photons keV$^{-1}$cm$^{-2}$s$^{-1}$) & (10$^{5}$ K) &
(km) & (10$^{5}$ K) & (m) & }
\startdata 
0.0--0.1 & A & 1.727 $\pm$ 0.068 & 6.42 $\pm$ 0.59 & (4.818) & 11.37 $\pm$ 0.35 &\nodata & \nodata & 1.128 (110)\\
        & B & 1.667 $\pm$ 0.183 & 6.92 $\pm$ 1.83 & (4.800) & 11.37 $\pm$ 0.36 & (17.1) & 0. $\pm$ 71. & 1.106 (109)\\
        & C & (1.684) & 6.65 $\pm$ 0.78  &  (4.800) & 11.72 $\pm$ 0.20 & (17.1) & 19. $\pm$ 20.  & 1.097 (110)\\
\hline\\
0.1--0.2 & A & 1.768 $\pm$ 0.101 & 5.14 $\pm$ 0.51 & (4.818) & 10.20 $\pm$ 0.32 &\nodata & \nodata & 1.129 (95)\\
        & B & 1.187 $\pm$ 0.195 & 3.27 $\pm$ 0.99 & (4.800) & 10.44 $\pm$ 0.32 & (17.1) & 79. $\pm$ 15. & 1.074 (94)\\
        & C & (1.684) & 4.30 $\pm$ 0.69  &  (4.800) & 10.68 $\pm$ 0.20 & (17.1) & 35. $\pm$ 13.  & 1.089 (95)\\
\hline\\
0.2--0.3 & A & 1.331 $\pm$ 0.092 & 2.94 $\pm$ 0.47 & (4.818) & 9.37 $\pm$ 0.28 &\nodata & \nodata & 1.373 (83)\\
        & B & 1.161 $\pm$ 0.196 & 2.64 $\pm$ 0.97 & (4.800) & 9.43 $\pm$ 0.29 & (17.1) & 27. $\pm$ 30. & 1.156 (82)\\
        & C & (1.684) & 3.52 $\pm$ 0.66  &  (4.800) & 9.64 $\pm$ 0.20 & (17.1) & 0. $\pm$ 31.  & 1.213 (83)\\
\hline\\
0.3--0.4 & A & 1.948 $\pm$ 0.114 & 5.04 $\pm$ 0.50 & (4.818) & 8.61 $\pm$ 0.31 &\nodata & \nodata & 1.379 (82)\\
        & B & 1.392 $\pm$ 0.268 & 3.53 $\pm$ 1.32 & (4.800) & 8.93 $\pm$ 0.28 & (17.1) & 71. $\pm$ 22. & 1.377 (81)\\
        & C & (1.684) & 3.85 $\pm$ 0.68  &  (4.800) & 9.09 $\pm$ 0.21 & (17.1) & 40. $\pm$ 11.  & 1.325 (82)\\
\hline\\
0.4--0.5 & A & 1.926 $\pm$ 0.066 & 6.97 $\pm$ 0.59 & (4.818) & 9.91 $\pm$ 0.33 &\nodata & \nodata & 0.952 (102)\\
        & B & 1.746 $\pm$ 0.184 & 6.96 $\pm$ 1.92 & (4.800) & 10.00 $\pm$ 0.33 & (17.1) & 42. $\pm$ 43. & 0.893 (101)\\
        & C & (1.684) & 6.26 $\pm$ 0.77  &  (4.800) & 10.23 $\pm$  0.21 & (17.1) & 46. $\pm$ 11.  & 0.892 (102)\\
\hline\\
0.5--0.6 & A & 2.081 $\pm$ 0.059 & 10.04 $\pm$ 0.67 & (4.818) & 10.53 $\pm$ 0.38 &\nodata & \nodata & 0.928 (120)\\
        & B & 1.770 $\pm$ 0.182 & 8.47 $\pm$ 2.26 & (4.800) & 10.64 $\pm$ 0.39 & (17.1) & 88. $\pm$ 32. & 0.930 (119)\\
        & C & (1.684) & 7.95 $\pm$ 0.84  &  (4.800) & 10.97 $\pm$  0.21 & (17.1) & 68. $\pm$ 9. & 0.911 (120)\\
\hline\\
0.6--0.7 & A & 1.936 $\pm$ 0.052 & 11.52 $\pm$ 0.71 & (4.818) & 10.95 $\pm$ 0.39 &\nodata & \nodata & 1.117 (131)\\
        & B & 1.791 $\pm$ 0.141 & 10.85 $\pm$ 2.30 & (4.800) & 11.05 $\pm$ 0.40 & (17.1) & 52. $\pm$ 44. & 1.108 (130)\\
        & C & (1.684) & 9.68 $\pm$ 0.90  &  (4.800) & 11.36 $\pm$ 0.21 & (17.1) &  59. $\pm$ 11. & 1.114 (131)\\
\hline\\
0.7--0.8 & A & 2.087 $\pm$ 0.063 & 9.59 $\pm$ 0.66 & (4.818) & 11.42 $\pm$ 0.41 &\nodata & \nodata & 1.260 (123)\\
        & B & 1.897 $\pm$ 0.184 & 7.95 $\pm$ 2.15 & (4.800) & 11.53 $\pm$ 0.40 & (17.1) & 69. $\pm$ 36. & 1.274 (122)\\
        & C & (1.684) & 7.31 $\pm$ 0.83 &  (4.800) & 11.86 $\pm$ 0.20 & (17.1) & 66. $\pm$ 9.  & 1.242 (123)\\
\hline\\
0.8--0.9 & A & 2.117 $\pm$ 0.080 & 6.53 $\pm$ 0.59 & (4.818) & 11.96 $\pm$ 0.42 &\nodata & \nodata & 1.203 (111)\\
        & B & 1.819 $\pm$ 0.270 & 6.32 $\pm$ 2.20 & (4.800) & 11.97 $\pm$ 0.42 & (17.1) & 50. $\pm$ 46. & 1.165 (110)\\
        & C & (1.684) & 4.99 $\pm$ 0.75  &  (4.800) & 12.37 $\pm$ 0.20 & (17.1) & 57. $\pm$ 10.  & 1.182 (111)\\
\hline\\
0.9--1.0 & A & 1.694 $\pm$ 0.070 & 6.72 $\pm$ 0.59 & (4.818) & 12.18 $\pm$ 0.39 &\nodata & \nodata & 1.317 (112)\\
        & B & 1.474 $\pm$ 0.156 & 5.67 $\pm$ 1.35 & (4.800) & 12.27 $\pm$ 0.19 & (17.1) & 69. $\pm$ 23. & 1.293 (111)\\
        & C & (1.684) & 6.64 $\pm$ 0.77  &  (4.800) & 12.54 $\pm$ 0.20 & (17.1) & 28. $\pm$ 17.  & 1.298 (112)\\
\enddata 
\tablecomments{ Values in parentheses are parameters that were frozen
  for the fits.}
\end{deluxetable}

\clearpage


\begin{thebibliography}{}
\bibitem[Bertch et al. (1992)]{be92} Bertch, D. L., et al. 1992,
    Nature 357, 306 
\bibitem[Bignami \& Caraveo (1996)]{bc96} Bignami, G. F., \& Caraveo,
  P. A. 1996, ARA\&A 34, 331
\bibitem[Caraveo et al. (1998)]{ca98} Caraveo, P. A., Lattanzi, M. G., Massone, G.,
    Mignani, R. P., Makarov, V.V., Perryman, M. A. C., \& Bignami, G. F. 1998,
    A\&A, 329, L1
\bibitem[Caraveo et al. (2004)]{ca04} Caraveo, P. A., De Luca, A.,
Mereghetti, S., Pellizzoni, A., \& Bignami, G. F. 2004,
    Science, 305, 376
\bibitem[De Luca et al. (2004)]{de04} De Luca, A., Caraveo, P. E.,
Mereghetti, S., Negroni, M., \& Bignami, G. F. 2004, \apj, 653, 1051
\bibitem[Fichtel et al. (1975)]{fi75} Fichtel, C. E. et al. 1975, \apj,
    198, 163
\bibitem[Halpern \& Holt (1992)]{hh92} Halpern, J. P., \&
    Holt, S. S. 1992, Nature, 357, 222
\bibitem[Halpern \& Ruderman (1993)]{hr93} Halpern, J. P., \& Ruderman, M.
    1993, \apj, 415, 286
\bibitem[Halpern \& Wang (1997)]{hw97} Halpern, J. P., \& Wang,
    F. Y.-H. 1997, \apj, 477, 905
\bibitem[Jackson et al.(2002)]{mj02} Jackson, M. S., Halpern, J. P.,
Gotthelf, E. V. \& Mattox, J. R. 2002 \apj, 578, 935
\bibitem[Kargaltsev et al. (2005)]{ka05} Kargaltsev, O. Y., Pavlov,
G. G., Zavlin, V. E., \& Romani, R. W. 2005, \apj, in press (astro-ph/0502076)
\bibitem[Mattox, Halpern, \& Caraveo (1998)]{ma98} Mattox, J. R.,
    Halpern, J. P., \& Caraveo, P. A. 1998, \apj, 493, 891
\bibitem[Mattox, Halpern, \& Caraveo (2000)]{ma00} Mattox, J. R.,
    Halpern, J. P., \& Caraveo, P. A. 2000, BAAS, 197, \#130.05
\bibitem[Shearer et al. (1998)]{sh98} Shearer, A., et al. 1998,
    A\&A, 335, 21
\bibitem[Thompson et al. (1977)]{th77} Thompson, D. J., Fichtel, C. E.,
    Hartman, R. C., Kniffen, D. A., \& Lamb, R. C. 1977, \apj, 213, 252
\end{thebibliography}
\end{document}